\documentclass[10pt]{article}
\usepackage[centertags]{amsmath}
\usepackage{amsfonts}
\usepackage{amssymb}
\usepackage{amsthm}
\usepackage{graphicx}
\usepackage{wrapfig}
\usepackage{ytableau}
\usepackage{pstricks}

\newcommand{\beq}{\begin{equation}}
\newcommand{\eeq}[1]{\label{#1}\end{equation}}
\newcommand{\bea}{\begin{eqnarray}}
\newcommand{\eea}[1]{\label{#1}\end{eqnarray}}

\newcommand{\eqn}{\begin{equation}}
\newcommand{\een}{\end{equation}}
\newcommand{\hV}{\hat{V}}
\newcommand{\hS}{\hat{S}}
\newcommand{\vf}{\varphi}
\newcommand{\nb}{\nabla}
\newcommand{\p}{\rho}
\newcommand{\m}{\mu}
\newcommand{\s}{\sigma}
\newcommand{\n}{\nu}
\newcommand{\ta}{\tau}

\newcommand{\BB}{\mathcal{B}}
\newcommand{\OO}{\mathcal{O}}
\newcommand{\AAc}{\mathcal{A}}
\newcommand{\CC}{\mathcal{C}}

\begin{document}
\setlength{\topmargin}{-1cm} \setlength{\oddsidemargin}{0cm}
\setlength{\evensidemargin}{0cm}

\begin{titlepage}
\begin{center}
{\Large \bf General Backgrounds for higher spin massive particles.}

\vspace{20pt}

{\large Ignacio Cortese$^a$ and Manuela Kulaxizi$^b$}

\vspace{12pt}
$^a$ Departamento de F\'isica de Altas Energ\'ias, Instituto de Ciencias Nucleares\\
Universidad Nacional Aut\'onoma de M\'exico,\\
Apartado Postal 70-543, Ciudad de M\'exico, 04510, M\'exico.\\
\vspace{6pt}
$^b$School of Mathematics, Trinity College Dublin, Dublin 2, Ireland.\\
\vspace{6pt}
nachoc@nucleares.unam.mx, manuela@maths.tcd.ie
\end{center}
\vspace{20pt}

\begin{abstract}
We consider the propagation of totally symmetric bosonic fields on generic background spacetimes. The mutual compatibility of the dynamical equations and constraints severely constrains the set of geometries where consistent propagation is possible. To enlarge this set in this article we allow several background fields to be turned on. 
We were able to show that massive fields of spin $s\geq 3$ may consistently propagate in a large set of non-trivial spacetimes, such as asymptotically de-Sitter, flat and anti-de-Sitter black holes geometries, as long as certain conditions between the various background fields are met. For the special case of massive spin-2 fields the set of allowed spacetimes is larger and includes doamin-wall-type geometries, such as the Freedman-Robertson-Walker metric. We comment on the assumptions underlying our study and on possible applications of our results.
\end{abstract}

\end{titlepage}

\newpage

\section{Introduction}

Consistent theories of interacting higher spin fields are difficult to construct. Higher spin gauge invariance poses strong constraints even in flat space, leading to several no-go theorems \cite{Weinberg:1964ew, Grisaru:1977kk, Grisaru:1976vm, Aragone:1979hx, Deser:1990bk, Weinberg:1980kq, Porrati:2008rm}. However, a non-linear, interacting theory has been succesfully constructed in AdS space \cite{Fradkin:1987ks, Vasiliev:1990en}. Moving away from AdS space still remains challenging.

In fact, even the free propagation of massive higher spin fields in non-trivial backgrounds is a challenging task. As first noted by Fierz and Pauli \cite{Fierz:1939ix}, the system of equations of motions and constraints governing the propagation of higher spin particles generically ceases to remain mutually compatible in the presence of nontrivial background fields. One may resort to a Lagrangian formulation, which is free from these kind of difficulties, but generically suffers from the Velo-Zwanziger problem, i.e., leads to superluminal propagation \cite{vz, Shamaly:1972zu, Hortacsu:1974bm, KobayashiShamaly, Deser:2000dz,Deser:2001dt}.

For a special class of spacetimes, those of constant curvature, a consistent Lagrangian desciption of free, massive higher spin fields is available as long as non-minimal terms are introduced \cite{Argyres:1989cu, Porrati:2010hm, Porrati:2009bs, Buchbinder:2012iz, Buchbinder:2000fy, Buchbinder:2007vq, Buchbinder:2007nq, BuchbinderKrykhtinLavrov, Zinoviev}. It is natural to wonder whether a consistent system of equations of motions and constraints can be found without the help of a Lagrangian formulation. Indeed this is possible even for backgrounds of non-constant curvature \cite{Cortese:2013lda, Kulaxizi:2014yxa}. 
The method employed in \cite{Cortese:2013lda} is rooted in a systematic deformation of the dynamical equations and constraints away from the flat case using the formalism developed in \cite{Kaparulin:2012px}.

In this article we will focus on totally symmetric, bosonic fields, of mass $m$ denoted by rank-s symmetric and traceless Lorentz tensors, {\it i.e.}, $\vf_{\mu_1\cdots\mu_s}$. The Fierz-Pauli equations describing the free propagation of these fields in flat space are given by:
\begin{equation}\label{fpflatzero}
(\partial^2-m^2)\vf_{\mu_1\cdots\mu_s}=0,\quad \partial \cdot \vf_{\mu_1\cdots\mu_{s-1}}=0,\quad \vf_{\mu_1\cdots\mu_{s-2}}'\equiv \vf_{\mu_1\cdots\mu_{s-1}^{\qquad \mu_{s-1}}}=0
\end{equation}
Here and in the following, the dot denotes the contraction of indices with respect to the relevant metric and the prime denotes the trace. The system is comprised of the Klein-Gordon equation and of the divergence/transversality and trace constraints. The constraints are necessary to reproduce the conrrect number of propagating degrees of freedom, given by $\binom{D-4+s}{s}+2 \binom{D-4+s}{s-1}$ in $D$ spactime dimensions.

The commuting nature of the ordinary derivatives renders the Fierz-Pauli system mutually compatible. This is no longer true in the presence of nontrivial background fields, because covariant derivatives do not commute with each other. Yet one may restore consistency by systematically including non-minimal terms to the covariantised version of (1) \cite{Kaparulin:2012px}. In practice, one requires the closure of the algebra generated by the deformed d'Alembertian, the divergence and the trace operators. The resulting set of equations of motion and constraints become mutually compatible, but there is a price to pay: the background fields are constrained to satisfy identities derived from the closure of the algebra. 

For the case of propagation in nontrivial spacetimes, \cite{Cortese:2013lda, Kulaxizi:2014yxa} showed that the following system allows for consistent propagation of higher spin fields 
\begin{align} 
&\left[\nb^2-m^2+\alpha R\right]\vf_{\m_1...\m_s}+s(s-1)R_{(\m_1}{}^{\n}{}_{\m_2}{}^{\p}
\vf_{\m_3\dots\m_s)\n\p}-sR_{\p(\m_1}\vf^{\p}{}_{\m_2\dots\m_s)}=0, \cr
&\nb\cdot\varphi_{\m_1\dots\m_{s-1}}=0,\quad \vf_{\mu_1\cdots\mu_{s-2}}=0
\end{align}
when $\alpha$ is the numerical constant
\eqn\label{alphadef}
\alpha={2(s-1)(s+D-2)\over (D-1)(D+2)},
\een
and the spacetime satisfies 
\eqn
Y_{\mu\nu\rho}=Z_{\mu\nu\rho}=0,\qquad X_{\mu\nu\rho}^{\quad\,\,\sigma\lambda}=0\,.
\een
Here $Y_{\mu\nu\rho},\,Z_{\mu\nu\rho},\,X_{\mu\nu\rho}^{\quad\,\,\sigma\lambda}$ are irreducible Lorentz tensors constructed out of the Ricci scalar, the Ricci curvature and the Weyl tensor of the background spacetime (explicit expressions can be found in eq. (\ref{XYZdef}). When $s=1$ there are no restrictions on the spacetime and when $s=2$, (4) reduces to $Y_{\mu\nu\rho}=Z_{\mu\nu\rho}=0$, leaving $X_{\mu\nu\rho}^{\quad\,\,\sigma\lambda}$ free to take any value.

Here we attempt to find a consistent set of dynamical equations and constraints for massive higher spins in more general spacetimes. We will use the same methodology, but we will allow additional background fields to be turned on. To simplify the analysis, we will focus on traceless, symmetric fields and consider deformations only to the dynamical equations and divergence constraints.

The organisation of this article is as follows: First we review the methodology and results of \cite{Kaparulin:2012px, Cortese:2013lda, Kulaxizi:2014yxa}. In section 3, we introduce further deformations to the involutive system by allowing more background fields to be turned on. By insisting on the vanishing of commutator between the d'Alembertian and the divergence, we determine necessary conditions the varioius background fields must meet. Our results are presented in section 4, where we also discuss background geometries of special interest.  In section 5 we detail the conditions necessary for consistent propagation to all orders in the curvature and we conclude in section 6, with a discussion of the methodology and its implicit assumptions together with possible applications of our results. Most of the technical details can be found in the appendices.

\paragraph{Conventions \& Notations:}

\noindent We use a mostly positive metric convention. The notation $a(n)$ denotes the collection of indices: $a_1\cdots a_n$. When the indices are placed in parenthesis, $(a_1\cdots a_n)$, the expression is totally symmetrised in all the indices $a_1,\cdots a_n$ with a normalisation factor ${1\over n!}$. 
The tensor symbol $g^{a_1\cdots a_n,\mu_1\cdots\mu_n}$, with $g^{\mu\nu}$ the inverse metric of the backrgound spacetime, denotes the following expression:
\eqn
g^{a_1\cdots a_n,\mu_1\cdots\mu_n}\equiv g^{a_1 b_1}\cdots g^{a_n b_n}\delta^{\mu_1\cdots\mu_n}_{b_1\cdots b_n}
\een
with $\delta^{\mu_1\cdots\mu_n}_{b_1\cdots b_n}=\delta^{\mu_1}_{b_1}\cdots\delta^{\mu_n}_{b_n}$.
The gravitational covariant derivative is denoted by $\nabla_\mu$ and the commutator obeys:
\eqn
[\nabla_\mu,\nabla_\nu] V^\rho=R^\rho_{\quad \sigma\mu\nu}\, V^\sigma\,.
\een
The notation $A^{a(n)}\cdot B_{b(m)}$ implies a signle index contraction as follows: 
\eqn
A^{a(n)}\cdot B_{b(m)}=A^{a(n)\mu} B_{\mu b(m)}\,.
\een

\section{Review of the formalism}

\noindent Consider the Fierz-Pauli system acting on symmetric and traceless higher spin fields:
\begin{equation}\label{fpflat}
T^{\mu_1\cdots\mu_s}\equiv (\partial^2-m^2)\vf^{\mu_1\cdots\mu_s}=0,\quad T^{\mu_1\cdots\mu_{s-1}}\equiv \partial \cdot \vf^{\mu_1\cdots\mu_{s-1}}=0\,, 
\end{equation}
This is an involutive system of partial differential equations of second order. The term involutive implies that the system comprises of all the differential consequences of order $\leq p$ from any order-p subsystem. 

From the differential consequences one can derive the so called ``gauge identities'', which should not be confused here with any gauge symmetries.  It is easy to see that the Fierz-Pauli system (\ref{fpflat}) satisfies the following ``gauge identity''
\eqn\label{giflat}
\partial_\mu T^{\,\,\mu\mu_1\cdots\mu_{s-1}}-(\partial^2-m^2) T^{\mu_1\cdots\mu_{s-1}}\equiv 0\,.
\een
The nature of the gauge identity can be made manifest if we express it as follows:
\eqn
[\partial_\mu, \partial^2-m^2] \vf^{\mu\mu(s-1)}=0\,; 
\een
it is clear that it corresponds to the vanishing of the commutator between the d'Alembertian operator and the divergence. Had we included the trace condition in the system, we would have had three gauge identities, in other words, three commuting operators defining an abelian algebra: the d'Alembertian, the divergence and the trace \cite{Rahman:2016tqc}. 

In what follows it will be useful to introduce the gauge identity generators: $L^{a_1\cdots a_{s-1}}$ 
 \eqn
 L^{a_1\cdots a_{s-1},\mu_1\cdots\mu_s}=\eta^{a_1\cdots a_{s-1},(\mu_1\cdots \mu_{}s-1}\partial^{\mu_s)},\quad L^{a_1\cdots a_{s-1},\mu_s\cdots \mu_{s-1}}=-\eta^{a_1\cdots a_{s-1},\mu_1\cdots \mu_{s-1}}(\partial^2-m^2)\,.
 \een
and express the gauge identity using the compact notation:
\eqn
L\triangleright T\equiv L_{\mu}^{\,\,a(s-1)} T^\mu=0\,,
\een
where the index $\mu$ collectively denotes the indices of each subsystem of the involutive system and the sum runs over all subsystems.

Now we consider coupling the Fierz-Pauli system to gravity and analyze the deformations order by order in the curvature. 
Starting from the minimal coupling ansatz, we promote all the derivatives into covariant ones: 
\begin{equation}
\begin{split}
T_0^{\mu_1\cdots\mu_s}&\equiv (\nabla^2-m^2)\vf^{\mu_1\cdots\mu_s}=0,\quad T_0^{\mu_1\cdots\mu_{s-1}}\equiv \nabla \cdot \vf^{\mu_1\cdots\mu_{s-1}}=0\,,\cr
 L_0^{a_1\cdots a_{s-1},\mu_1\cdots\mu_s}&=g^{a_1\cdots a_{s-1},(\mu_1\cdots \mu_{}s-1}\nabla^{\mu_s)},\quad L_0^{a_1\cdots a_{s-1},\mu_s\cdots \mu_{s-1}}=-g^{a_1\cdots a_{s-1},\mu_1\cdots \mu_{s-1}}(\nabla^2-m^2)\,.
\end{split}
\end{equation}
The resulting system is consistent to zeroth-order in the curvature. At linear order the covariantised gauge identity (7) fails to be satisfied due to the non-commutativity of the covariant derivatives. This is precisely the anomaly noticed by Fierz and Pauli \cite{Fierz:1939ix}:
\begin{align}\label{anomalya}
\AAc^{a(s-1)}\equiv -{(L_0\triangleright T_0)^{\mu(s-1)}} =[\nabla^2,\nabla_\mu]\,\vf^{\mu \, a(s-1)} =\OO(R)
\end{align}
To push the failure of the gauge identities to $\OO(R^2)$ we introduce first order deformations to the dynamical equations, $T_1$, and gauge identity generators, $L_1$:
\eqn\label{eqanomaly}
(L_0+L_1)\triangleright (T_0+T_1)=L_0\triangleright T_0+L_1\triangleright T_0+L_0\triangleright T_1+L_1\triangleright T_1=\OO(R^2)\,.
\een
This implies that the first order deformations should satisfy
\eqn\label{giaone}
L_0\triangleright T_1+L_1\triangleright T_0 +\OO(R^2)=-L_0\triangleright T_0 +\OO(R^2)=[\nabla^2,\nabla_\mu]\,\vf^{\mu\, a(s-1)} +\OO(R^2)
\een
In practice, one tries to bring the expression for the anomaly term (\ref{anomalya}) in a form which will allow for the identification of the apporpriate deformations $L_1$ and $T_1$ in accordance with (\ref{giaone}).
The proposed identification may sometimes give non-local solutions for the first order deformations. If we insist on preserving locality and the number of degrees of freedom, we may have to impose restrictions on the external background. Constraints on the higher spin field are not allowed since they will ruin the involutive form of the system and/or the number of degrees of freedom count.

As shown in \cite{Cortese:2013lda, Kulaxizi:2014yxa} the anomaly tensor for the covariantised Fierz-Pauli system can be expressed as follows:
\eqn\label{anomalyb}
\begin{split}
&\AAc^{a (s-1)}=\nabla_\mu \left[s(s-1)\, R^{(\mu\quad a_1}_{\nu\quad \rho}\vf^{a_2\cdots a_{s-1})\nu\rho}-s R^{(\rho(\mu}\vf_\rho^{\,\,a_1\cdots a_{s-1)}}-\alpha R\vf^{a_1\cdots a_{s-1}\mu}\right]-\cr
&-(s-1)\left[ (s-2) R^{(a_1\quad a_2}_{\mu\quad\nu}\nabla\cdot \vf^{a_3\cdots a_{s-1})\mu\nu}-R^{\mu(a_1}\nabla\cdot\vf_\mu^{a_2\cdots a_{s-1)}} +{\alpha\over s-1} R\,\nabla\cdot \vf^{a(s-1)}    \right]+\BB^{a(s-1)}\,,
\end{split}\een
where $R_{\mu\nu\rho\sigma},\,R_{\mu\nu},R$ respectively denote the Riemann curvature tensor, the Ricci tensor and the Ricci scalar of the external spacetime, and $\alpha$ is defined in (\ref{alphadef}). Comparing eq. (\ref{giaone}) with the first order terms of the gauge identity (\ref{anomalyb}) we identify the first order deformation of the dynamical equationsfrom the first line of (\ref{anomalyb}):
\eqn\label{Tonedef}
T_1^{\mu_1\cdots\mu_s}=s(s-1) R^{(\mu_1\quad \mu_2}_{\mu\quad\rho} \vf^{\mu_3\cdots\mu_s)\nu\rho }-s R^{\rho(\mu_1}\vf_\rho^{\mu_2\cdots\mu_s}-\alpha R\vf^{a_1\cdots a_{s-1}\mu} ,\quad T_1^{\mu_1\cdots\mu_{s-1}}=0\,,
\een
and the first order deformation of the gauge identity generators from the second line of (\ref{anomalyb}):
\eqn\label{Lonedef}
\begin{split}
&L^{a(s-1)}_{1\quad\quad\mu (s-1)}=-(s-1)\left[(s-2) \delta^{a_1\cdots a_{s-1}}_{\rho\sigma(\mu_1\cdots\mu_{s-3}}R^{\rho\quad\,\,\sigma}_{\,\,\, \mu_{s-2}\,\,\,\mu_{s-1})}-\delta^{a_1\cdots a_{s-1}}_{\rho(\mu_1\cdots\mu_{s-2}} R^\rho_{\,\, \mu_{s-1})}+{\alpha\over s-1} \delta^{a_1\cdots a_{s-1}}_{(\mu_1\cdots\mu_{s-1})} R\right],\cr
&L^{a(s-1)}_{1\quad\quad\mu (s)}=0\,\,.
\end{split}
\een
The remaining terms denoted collectively as $\BB^{a(s-1)}$ must be set to zero.  $\BB^{a(s-1)}$ contains gradients of the Riemann and Ricci tensors which can be split into irreducible Lorentz tensors resulting in \cite{Cortese:2013lda, Kulaxizi:2014yxa}:
\begin{align}\label{BBLL}
\mathcal{B}^{a(s-1)}=- \frac{(s-1)(s-2)}{D-2} \left[(d-2) X_{\mu\nu\rho}^{\qquad(a_1a_2}\varphi^{a_3\cdots a_{s-1})\mu\nu\rho}+Y_{\mu\nu\rho} \,g^{(a_1 a_2}\varphi^{a_3\cdots a_{s-1})\mu\nu\rho} \right]\nonumber\\
+\left(\frac{s-1}{D-2} \right)\left[(D-6+2 s) Y_{\mu\nu}^{\quad(a_1}\varphi^{a_2\cdots a_{s-1})\mu\nu}-\left(\frac{s+2D-6}{3}\right)Z_{\mu\nu}^{\quad(a_1}\varphi^{a_2\cdots a_{s-1})\mu\nu}\right]\,.\nonumber\\
\end{align}
where
\eqn\label{XYZdef}
\begin{split}
X_{\mu\nu\rho}^{\qquad\alpha\beta}&=\nabla_{(\mu}\CC_{\nu\quad\rho )}^{\alpha\quad\beta}-\left({2\over D+2}\right) g_{(\mu\nu}\nabla^\sigma \CC_{\rho)\quad\sigma}^{(\alpha\quad\beta)}\cr
Y_{\mu\nu\rho}&=\nabla_{(\mu}R_{\nu\rho)}-\left({2\over D+2}\right) g_{(\mu\nu}\nabla_{\rho)}R\cr
Z_{\mu\nu\rho}&=2 \nabla_{[\rho}R_{\mu]\nu}+\left({1\over D-1}\right) g_{\nu[\rho}\nabla_{\mu]}R+(\mu\leftrightarrow \nu)=2 {D-2\over D-3} \nabla^\sigma\, \CC_{\rho (\mu\nu)\sigma}\,,
\end{split}
\een
with $\CC_{\mu\nu\alpha\beta}$ the Weyl tensor.

Summarising, as long as $\BB^{a(s-1)}=0$ or equivalently, $Y_{\mu\nu\rho}=Z_{\mu\nu\rho}=X_{\mu\nu\rho}^{\quad a_1a_2}=0$ the involutive system (2) is consistent to first order in the curvature. It is however easy to see that given the above restrictions on the background spacetime, (2) is consistent to all orders in the curvature. Firstly, the covariantised anomaly tensor (\ref{anomalya}) does not contain $\OO(R^2)$ terms and secondly, all second and higher order terms in (\ref{eqanomaly}), {\it i.e.}, $L_1\triangleright T_1$, vanish by virtue of (\ref{Tonedef}) and (\ref{Lonedef}).

\section{Deformations including more general backgrounds.}

\noindent As reviewed above, deformations $T_1$ of the covariantised Fierz-Pauli system $T_0$ are allowed in background spacetimes fulfilling
\begin{equation}
\label{badt}
  \mathcal{B}^{a(s-1)}=0\,.
\end{equation}
This condition is quite strong and consistent propagation is achieved for a very limited set of nontrivial backgrounds, such as constant curvature spaces and their products, including a particular set of domain-wall geometries \cite{Cortese:2013lda, Kulaxizi:2014yxa}. Here we would like to investigate the possibility of enlarging the allowed set of spacetimes upon which the deformed involutive system of Fierz and Pauli is mutually compatible. In particular, we are interested in exploring this scenario by turning on more background fields. In practice, we would like to ask the question: are there any further modifications to $T=T_0+T_1$ allowing for more general backgrounds where higher spin particles can propagate consistently? In other words, is it possible to find deformations which would respect the gauge identity?

Let us consider deformations of the system $T=T_0+\widetilde{T}_1$ in the form
\begin{subequations}\label{gendef}
  \begin{align}
  \widetilde{T}_1^{\mu(s)} &= T_1^{\mu(s)} +Q_1^{\mu(s)} \label{gendefEoM}\\
  \widetilde{T}_1^{\mu(s-1)} &= T_1^{\mu(s-1)} + Q_1^{\mu(s-1)}, \label{gendefCon}
\end{align}
\end{subequations}
with $T_1^{\mu(s)}$ and $T_1^{\mu(s-1)}$ given in eq. (\ref{Tonedef}).
The $Q_1^{\mu(s)}$ and $Q_1^{\mu(s-1)}$ tensors should be completely symmetric and traceless and composed of background fields other than the curvature. In writing (\ref{gendef}) we are implicitly making the assumption that consistent propagation is always possible in constant curvature spaces (implicit in using $T_1$ as defined in (\ref{Tonedef})) \footnote{We thank R. Rahman for clarifying this assumption to us.}. On the other hand, we remain agnostic as to the specific background field content of the tensors $Q_1$; we simply demand that there exist some other background fields such that $Q_1^{\mu(s)},\, Q_1^{\mu(s-1)}$ are of the same order as the curvature. For reasons of simplification, we also preserve the structure of the covariant derivative. This is a very strong assumption; we know for example that in string theory, higher spin fields are charged under several background fields other than the metric. We hope to lift this assumption in future work.

The deformations (\ref{gendef}) will in principle be accompanied by deformations of the gauge identity generators:
\begin{subequations}\label{gaugeiddef}
\begin{align}
\widetilde{L}_1=L_1+M_1
\end{align}
\end{subequations}
with the $L_1$ given in (\ref{Lonedef}) and $M_1$ to be specified by consistency requirements. Demanding that the gauge identity is satisfied to first order in the curvature leads to:
\begin{equation}\label{gaugeid}
  \stackrel{=-\mathcal{A}^{a(s-1)}}{\overbrace{(L_0\triangleright T_0)^{a(s-1)}}} + (L_0\triangleright Q_1)^{a(s-1)}+\stackrel{=\text\tiny{eq. (\ref{giaone})} }{\overbrace{(L_0\triangleright T_1)^{a(s-1)}+(L_1\triangleright T_0)^{a(s-1)}}}+(M_1\triangleright T_0)^{a(s-1)}=\OO(R^2)\,.
\end{equation}
Recall that $T_1,\,L_1$ are defined so that:
\begin{equation*}
 -L_0\triangleright T_0={L_0}^{a(s-1)}{}_{\mu(s)}{T_1}^{\mu(s)}+
  {L_1}^{a(s-1)}{}_{\mu(s-1)}{T_0}^{\mu(s-1)}+\mathcal{B}^{a(s-1)}\,.
\end{equation*}
with $\BB^{a(s-1)}$ given in eq.(\ref{BBLL}).
Hence, the vanishing of the gauge identity to first order in the curvature yields:
\eqn\label{cure}
  \mathcal{B}^{a(s-1)}= {L_0}^{a(s-1)}{}_{\mu(s)}{Q_1}^{\mu(s)}
  + {L_0}^{a(s-1)}{}_{\mu(s-1)}{Q_1}^{\mu(s-1)}+{M_1}^{a(s-1)}{}_{\mu(s)} {T_0}^{\mu(s)}+{M_1}^{a(s-1)}{}_{\mu(s-1)} {T_0}^{\mu(s-1)},\,.
\een
Our task will be to specify  $Q_1^{\mu(s)}, Q_1^{\mu(s-1)}$ and $M_1^{a(s-1)}{}_{\mu(s)},M_1^{a(s-1)}{}_{\mu(s)}$ such that (\ref{cure}) is true.

Note that the terms involving $M_1$ should cancel certain $\OO(R)$ terms coming from the first line of (\ref{cure}). To this order, they do not have an impact on the consistency conditions of the background fields. However, they generically produce new, non-trivial $\mathcal{O}(R^2)$ terms on the right-hand side of equation (\ref{gaugeid}). Requiring then that (\ref{gaugeid}) is true to all orders in the deformation, will result in additional conditions for the background fields which are explored in section 5.

Let us now discuss in detail the contributions to the deformed Fierz-Pauli equations coming from the background fields. Generic tensors $Q^{\mu(s)}$ and $Q^{\mu(s-1)}$ with the required symmetry properties are constructed from all the possible couplings between irreducible tensors of mixed symmetry and the higher spin fields:
\begin{equation}\label{Utensor}
\begin{split}
  Q_1^{\mu(s)} = &\ \ a_1 U \varphi^{\mu(s)} + a_2 U_\rho{}^{(\mu_1} \varphi^{\mu_2\dots\mu_s)\rho} \\
   & +a_3 \left( V_\rho{}^{(\mu_1} \varphi^{\mu_2\dots\mu_s)\rho} - \frac{s-1}{D-4+2s}V_{\rho\sigma}g^{(\mu_1\mu_2}\varphi^{\mu_3\dots\mu_s)\rho\sigma} \right) \\
   & +\left(a_4\hat{V}_{\rho | \sigma}{}^{(\mu_1\mu_2} + a_5 \hat{V}^{(\mu_1 | \mu_2}{}_{\rho\sigma}  \right) \varphi^{\mu_3\dots\mu_s)\rho\sigma} \\
   & -\frac{s-2}{D-4+2s}\left[ (2a_4 +a_5)\hat{V}_{\rho | \sigma\nu}{}^{(\mu_1} + a_5 \hat{V}^{(\mu_1 |}{}_{\rho\sigma\nu} \right]g^{(\mu_2\mu_3}\varphi^{\mu_4\dots\mu_s)\rho\sigma\nu} \\
   & + a_6 \bigl[ V_{\rho\sigma}{}^{(\mu_1\mu_2}\varphi^{\mu_3\dots\mu_s)\rho\sigma} - \frac{s-2}{D-4+2s}\bigl( 2V_{\rho\sigma\nu}{}^{(\mu_1}g^{\mu_2\mu_3}\varphi^{\mu_4\dots\mu_s)\rho\sigma\nu} \\
   &-\frac{s-3}{D-6+2s}V_{\rho\sigma\nu\lambda}g^{(\mu_1\mu_2}g^{\mu_3\mu_4}\varphi^{\mu_5\dots\mu_s)\rho\sigma\nu\lambda}
   \bigr)\bigr] \\
   & +a_7 W_\rho{}^{(\mu_1}{}_\sigma{}^{\mu_2}\varphi^{\mu_3\dots\mu_s)\rho\sigma},
\end{split}
\end{equation}
and
\begin{equation}\label{Stensor}
\begin{split}
  Q_1^{\mu(s-1)} = &\ \ b_1 S_\rho \varphi^{\mu(s-1)\rho} \\
   & + b_2 \left( S_{\rho\sigma}{}^{(\mu_1}\varphi^{\mu_2\dots\mu_{s-1})\rho\sigma} - \frac{s-2}{D-6+2s}S_{\rho\sigma\nu}g^{(\mu_1\mu_2}\varphi^{\mu_3\dots\mu_{s-1})\rho\sigma\nu} \right) \\
   & + b_3 \hat{S}^{(\mu_1}{}_{\rho\sigma}\varphi^{\mu_2\dots\mu_{s-1})\rho\sigma}.
   \end{split}
\end{equation}
where $(a_1,\cdots a_7)$ and $(b_1,\cdots b_3)$ are arbitrary coefficients. All the information on the external background fields is hidden in the irreducible tensors of (\ref{Utensor}) and (\ref{Stensor}) which correspond to the following Young diagrams :
\begin{eqnarray*}
\ytableausetup{smalltableaux}
  U \thicksim \circ \ \ & U_{\rho\sigma} \thicksim
\begin{ytableau}
~ \\
~
\end{ytableau} \ \ & V_{\rho\sigma} \thicksim
\begin{ytableau}
~ & ~
\end{ytableau} \ \  \\
  \hat{V}_{\rho | \sigma\mu\nu} \thicksim
\begin{ytableau}
~ & ~ & ~ \\
~
\end{ytableau} \ \ & V_{\rho\sigma\nu\lambda} \thicksim
\begin{ytableau}
~ & ~ & ~ & ~
\end{ytableau} \ \ & W_{\rho\sigma\nu\lambda} \thicksim
\begin{ytableau}
~ & ~ \\
~ & ~
\end{ytableau}
\end{eqnarray*}
Tensors corresponding to the diagrams
\begin{equation*}
  U_{\rho\sigma\nu\lambda} \thicksim
\begin{ytableau}
~ \\
~ \\
~ \\
~
\end{ytableau}\quad \text{and} \quad  \hat{U}_{\rho\sigma\nu | \lambda} \thicksim
\begin{ytableau}
~ & ~ \\
~ \\
~
\end{ytableau}
\end{equation*}
do not couple to the higher spind fields due to symmetry properties. Similarly, the deformation of the constraint equation involves coupling the fields to the tensors:
\begin{equation*}
  S_\rho \thicksim
\begin{ytableau}
~
\end{ytableau} \ \  S_{\rho\sigma\nu} \thicksim
\begin{ytableau}
~ & ~ & ~
\end{ytableau} \ \   \hat{S}_{\rho | \sigma\nu} \thicksim
\begin{ytableau}
~ & ~ \\
~
\end{ytableau}
\end{equation*}
Tensors in the following representations 
\begin{equation*}
  S \thicksim \circ \qquad  P_{\rho\sigma\nu} \thicksim
\begin{ytableau}
~ \\
~ \\
~
\end{ytableau}.
\end{equation*}
do not couple to the higher spin fields. Finally, higher rank background tensors are not expected to contribute in the anomaly cancellation.

In order to compute the gauge identity we need to perform basically the computations
\begin{equation}\label{gidpieces}
  \nabla_{\mu_s}{Q_1}^{\mu(s)} \quad  \text{and} \quad  -(\nabla^2-m^2){Q_1}^{\mu(s-1)},
\end{equation}
involved in the $(L_0 \triangleright Q_1)^{a(s-1)}$ terms of eq. (\ref{cure}). In practice, we need to bring the sum of the two terms in (\ref{gidpieces}) in the form $-M_1\triangleright T_0+ rest$ by pushing the derivatives all the way to the higher spin fields. The computations are tedious but straightforward. Here we will give a few examples, for more details the interested reader should consult appendix B. 

The first term in $Q_1^{\mu(s)}$ is given by $a_1 U\vf^{\mu(s)}$. We need to compute:
\eqn\label{ega}
a_1\nabla_\nu (U\vf^{\mu(s-1)\nu})=a_1(\nabla_\nu U) \vf^{\mu(s-1)\nu}+a_1 U\stackrel{T_0^{\mu(s-1)}} {\overbrace{ (\nabla\cdot\vf^{\mu(s-1)} )} }\,.
\een
The second term in (\ref{ega}) is exactly of the type we are after, defining a part of $M_1$. The first term must be cancelled against terms in $\BB^{\mu(s-1)}$ coming from the orginal anomaly.
Let us do one more example, which is slightly less trivial:
\eqn\label{egb}
\begin{split}
&a_2 \nabla_\nu \left(U_\rho^{\,\,\,\, ( \mu_1}\vf^{\mu_2\cdots\mu_{s-1} \nu)\rho}\right)={a_2\over s}\nabla_\nu\left[U_\rho^{\,\,\,\,\nu} \vf^{\mu(s-1)\rho}+\sum_{n=1}^{s-1}U_\rho^{\,\,\, \mu_n}\vf^{\mu_1\cdots\mu_{n-1}\mu_{n+1}\cdots\mu(s-1)\nu\rho}\right]=\cr
&={a_2\over s}\left[(-\nabla\cdot U_\rho)\vf^{\mu(s-1)\rho}+U_{\rho\nu}\nabla^\nu\vf^{\mu(s-1)\rho}+\sum_{n=1}^{s-1}\left(\nabla_\nu U_\rho^{\,\,\,\, \mu_n}\vf^{\mu_1\cdots\mu_{n-1}\mu_{n+1}\cdots\mu_{s-1}\nu\rho}+U_\rho^{\,\,\,\, \mu_n}(\nabla\cdot\vf)^{\mu_1\cdots\mu_{n-1}\mu_{n+1}\cdots\mu_{s-1}\rho}\right)\right]=\cr
&={a_2\over s}\left[(-\nabla\cdot U_\rho)\vf^{\mu(s-1)\rho}-U_{\nu\rho}\nabla^{[\nu}\vf^{\rho]\mu(s-1)} +(s-1)\left(\nabla_{(\nu}U_{\rho)}^{\,\,\,\, (\mu_1}\vf^{\mu_2\cdots\mu_{s-1)} )\nu\rho}+U_\rho^{\,\,\,\,(\mu_1} \stackrel{T_0^{\mu(s-1)}} {\overbrace{ \nabla\cdot\vf^{\mu_2\cdots\mu_{s-1})\rho} }}\right)\right] \,,
\end{split}
\een
where we used the definition of indices symmetrisation to go from the first to second line and from the second to the third. Once more there are terms which cannot be written in the form $(M_1  \triangleright T_0 )^{a(s-1)}$ and should be cancelled by terms in the anomaly. With similar techniques we can work out all the relavant terms produced by hitting with a derivative on $Q_1^{\mu(s)}$ of eq.(\ref{Utensor}). Likewise we treat terms of the form $ -(\nabla^2-m^2){Q_1}^{\mu(s-1)}$. Computations in this case are slightly simpler. As an example consider:
\eqn
\begin{split}
b_2(\nabla^2-m^2) \left(S_{\rho\sigma}^{\,\,\,(\mu_1}\vf^{\mu_2\cdots\mu_{s-1})\rho\sigma}\right)&=b_2\left(\nabla^2S_{\rho\sigma}^{\,\,\,(\mu_1}\right)\vf^{\mu_2\cdots\mu_{s-1})\rho\sigma}+2\,b_2 \left(\nabla^\nu S_{\rho\sigma}^{\,\,\,(\mu_1}\right)\nabla_\nu \vf^{\mu_2\cdots\mu_{s-1})\rho\sigma}+\cr
&+b_2 \, S_{\rho\sigma}^{\,\,\,(\mu_1} \stackrel{T_0^{\mu(s)}} {\overbrace{ (\nabla^2-m^2) \vf^{\mu_2\cdots\mu_{s-1})\rho\sigma}  }} 
\end{split}
\een
Proceeding this way leads to:
\eqn
(L_0\triangleright Q_1)^{a(s-1)}=-(M_1\triangleright T_0)^{a(s-1)}+\CC^{a(s-1)},\,
\een
where $\CC^{a(s-1)}$ is a complicated expression involving the extra background fields and their derivatives, parts of which we saw in the examples above. The explicit form of $M_1$ can be found in Appendix D. Preservation of the gauge identity to first order in the curvature demands that:
\eqn\label{con}
\CC^{a(s-1)}=\BB^{a(s-1)}\,.
\een
This condition corresponds to a system of differential equations the background fields should satisfy for higher spin fields to porpagate consistently on them. Explicit results are presented in the following section. The reader interested in the intermediate steps of the computation is encouraged to consult Appendices B and C.

\section{Results}

\noindent In this section we focus on the conditions the backgrounds fields should satisfy. For fields of generic spin $s\geq 3$, eq (\ref{con}) allows for a very small set of non-vanishing background tensors, namely:
\eqn\label{bgnonzero}
U,U_{\mu\nu}, W_{\mu\nu\rho\sigma}, S_\nu,\hat{S}_{\mu|\rho\sigma}, Y_{\mu\nu\rho}, Z_{\mu\nu\rho}, X_{\mu\nu\rho}^{\qquad a_1a_2}\,.
\een
In other words, without loss of generality one may set\footnote{With the help of the identities in Appendix B, one can show that all the allowed terms in (\ref{Utensor}) are proportional to $(a_4-a_5)$ and thus vanish.}:
\eqn
b_2=0=a_3=a_4-a_5=a_6=0\,.
\een
The remaining non-trivial background fields (\ref{bgnonzero}) must satsify the following set of equations
\eqn\label{cona}
\begin{split}
X_{\nu\rho\sigma}^{\quad \mu_1\mu_2}&=-{a_7\over s(s-1)}\nabla_{(\nu} W_{\rho\quad \sigma )}^{\,\,\, \mu_1\quad \mu_2}\cr
Y_{\mu\nu\rho}&=0\cr
Z_{\nu\rho\mu}&=-\left({3(D-2)\over 2D-6+s}\right) \left({a_2\over s}\right)\, \nabla_{(\nu}U_{\rho)\mu},  \cr
\end{split}
\een
and\
\eqn\label{conaa}
\begin{split}
{a_2\over s}U_{\nu\rho}&=-2 b_1\nabla_\nu S_\rho,\cr
a_1\nabla_\nu U=\nabla^\rho U_{\rho\mu}=0=\nabla_{(\nu}S_{\rho )}&=\nabla^\rho S_\rho= \nabla^2 S_\rho,\cr
{a_7\over s} W^{(\rho\quad \sigma)}_{\quad \nu\quad \mu}&=b_3\nabla_\nu \hS_{\mu |}^{\quad \rho\sigma},\cr
\nabla^\rho W_{\rho\quad \sigma}^{\,\,(\mu_1\quad \mu_2)}=0=\nabla^\nu \hS_{\nu|\rho\sigma}=\nabla^\rho \hS_{\nu|\rho\sigma}&=\nabla^2 \hS_{\nu|\rho\sigma}=\nabla_{[\nu}\hS_{\mu]|\rho\sigma}\,.
\end{split}
\een
The latter conditions, (\ref{conaa}) follow from the tracelessness and the other symmetry properties of the fields involved in (\ref{cona}).
Assuming the above equations are satisfied, propagation of higher spin fields is described by the following involutive system\footnote{Notice that background scalar function $U$ is allowed as long as it satisfies : $a_1\nabla_\nu U=0$. So $U$ is a constant. This reflects the ambiguity in the definition of the mass parameter in arbitrary spacetimes (recall that U couples to the higher spin field exactly like the mass). The simplest solution is to set: $a_1=0$}:
\eqn\label{eqnfinals}
\begin{split} 
&\left[\nb^2-m^2+\alpha R+a_1 U\right]\vf_{\m_1...\m_s}+s(s-1)R_{(\m_1}{}^{\n}{}_{\m_2}{}^{\p}
\vf_{\m_3\dots\m_s)\n\p}-sR_{\p(\m_1}\vf^{\p}{}_{\m_2\dots\m_s)}+\cr
&\qquad\qquad\qquad\qquad\qquad\qquad\qquad +a_2 U_\p^{\,\,\,(\m_1}\vf^{\,\m_2\cdots\m_s )\p}+ a_7 W_\rho{}^{(\mu_1}{}_\sigma{}^{\mu_2}\varphi^{\mu_3\dots\mu_s)\rho\sigma}   =0,\cr
&\nb\cdot\varphi_{\m_1\dots\m_{s-1}}+b_1 S_\p \vf^{\m_1\cdots\m_{s-1}\rho}+b_3 \hS^{(\m_1}_{\,\,\,\p\s}\vf^{\,\m_2\cdots\m_{s-1})\p\s}=0,\qquad \vf^{\mu(s-2)}=0
\end{split}
\een
with $\alpha$ the numerical constant
\eqn
\alpha={2(s-1)(s+D-2)\over (D-1)(D+2)}\,.
\een

The case of spin-2 fields is special. As expected, there is more freedom in the choice of background fields. In particular, $X_{\nu\rho\sigma}^{\quad\mu_1\mu_2}$ is not constrained and the same is true for several of the additional background fields introduced. For a {\it minimal} solution we set:
\eqn
a_4-a_5=a_6=a_7=0\,.
\een
The remaining background fields should satisfy:
\eqn\label{conb}
\begin{split}
{a_2\over 2}U_{\nu\rho}&=-2 b_1 \nabla_\nu S_\rho\cr
a_1\nabla_\nu U=0=-b_1\nabla^2 S_\rho={a_2\over 4}\nabla^\nu U_{\nu\rho},&\quad a_3=0\cr
{a_2 \over 2}\nabla_{(\nu}U_{\rho)\mu}+b_2\nabla^2 S_{\nu\rho\mu}+b_3\nabla^2 \hS_{\mu|\nu\rho}&=Y_{\mu\nu\rho}-{2\over 3} Z_{\nu\rho\mu}\,,
\end{split}
\een
and the involutive system reduces to
\eqn\label{eqnfinalstwo}
\begin{split} 
&\left[\nb^2-m^2+\alpha R+a_1 U\right]\vf_{\m_1...\m_s}+2 R_{(\m_1}{}^{\n}{}_{\m_2}{}^{\p}
\vf_{\m_3\dots\m_s)\n\p}-2 R_{\p(\m_1}\vf^{\p}{}_{\m_2\dots\m_s)}+a_2 U_\p^{\,\,\,(\m_1}\vf^{\,\m_2\cdots\m_s )\p}=0,\cr
&\nb\cdot\varphi_{\m_1\dots\m_{s-1}}+b_1 S_\p \vf^{\m_1\cdots\m_{s-1}\rho}+b_2 S_{\rho\sigma}^{\,\,\,(\mu_1}\vf^{\mu_2\cdots\mu_{s-1})\rho\sigma}+b_3 \hS^{(\m_1}_{\,\,\,\p\s}\vf^{\,\m_2\cdots\m_{s-1})\p\s}=0,\qquad\vf^{\mu(s-2)}=0
\end{split}
\een
with $\alpha$ the numerical constant
\eqn
\alpha={2D\over (D-1)(D+2)}\,.
\een

Let us end this section with a final comment. Eqs (\ref{cona},\,\ref{conaa}) and (\ref{conb}) should be viewed as (part of) the set of differential equations imposed upon the background fields by an underlying consistent, but otherwise unkown theory, where massive higher spin fields can propagate.

\subsection{General comments on the solutions}

\paragraph{Fields of spin $s\geq 3$.}

\begin{itemize}

\item The first observation we can make is that spacetimes with non-vanishing $Y_{\mu\nu\rho}$ are not allowed even in the presence of other background fields. This is a very strong constraint and excludes all conformally flat spacetimes except for the ones studied in \cite{Kulaxizi:2014yxa}.

\item Given a background geometry, one can try to solve the conditions (\ref{cona}) to obtain explicit expressions for the other background fields in terms of the geometric data. In general, this is a difficult problem. We can simplify it by noticing the following consequences of (\ref{cona}) and (\ref{conaa}):
\eqn\label{conaasa}
\begin{split}
R_{\lambda\nu}S^\lambda&=0, 
\end{split}
\een
and
\eqn\label{conaasb}
\begin{split}
R_{\lambda\nu}\hS^{\lambda|}_{\quad \rho\sigma}-&2\,\, \hS^{\mu|}_{\quad\lambda(\sigma}R^\lambda_{\,\,\rho)\mu\nu}=0\qquad \Rightarrow\qquad R_{\kappa\sigma\lambda\rho}S^{\lambda|\rho\kappa}=0\cr
\end{split}
\een
Notice that (\ref{conaasa}) and (\ref{conaasb}) define an algebraic system of $n$ equations for $n$ unkowns; where $n$ are the independent components of the unkown background fields $S_\rho, S_{\lambda|\rho\sigma}$\footnote{In the former case, $n=D$ whereas in the latter $n={D(D-1)\over 2}$}. Thus, given a certain background spacetime, $S_\rho$ and $\hS_{\mu|\rho\lambda}$ can be completely determined from (\ref{conaasa}) and (\ref{conaasb}). Recall that knowledge of $(S_\rho\, \hS_{\mu|\rho\lambda})$ is sufficient to completely determine the involutive system since:
\eqn
\begin{split}
Z_{\nu\rho\mu}=&-\left({3(D-2)\over 2D-6+s}\right) \left({a_2\over s}\right)\, \nabla_{(\nu}U_{\rho)\mu}\qquad\text{and}\qquad {a_2\over s} U_{\nu\rho}=-2 b_1\nabla_{[\nu} S_{\rho]}\cr
X_{\nu\rho\sigma}^{\qquad \mu_1\mu_2}=&-{a_7\over s(s-1)}\nabla_{(\nu} W_{\rho\quad \sigma )}^{\,\,\, \mu_1\quad \mu_2},\qquad\qquad\quad\text{and}\qquad {a_7\over s} W^{(\rho\quad \sigma)}_{\quad \nu\quad \mu}=b_3\nabla_\nu \hS_{\mu |}^{\quad \rho\sigma}\,.
\end{split}
\een
Hence finding an appropriate solution has been reduced to a simple algebraic problem with the use of (\ref{conaasa}) and (\ref{conaasb}).
Once a solution is determined, it should be checked that (\ref{cona}) and (\ref{conaa}) are identically satisfied. If this is not the case, then a solution does not exist.

Notable cases of background geometries which may be interesting to study are those where besides $Y_{\mu\nu\rho}=0$, $X_{\m\n\p}^{\qquad \s\ta}$ or $Z_{\m\n\p}$ also identically vanish. 
\begin{itemize}
\item When $Y_{\mu\nu\rho}=X_{\mu\nu\rho}^{\qquad\mu_1\mu_2}=0$, the background spacetime should satisfy:
\eqn
\nabla_{(\mu}\CC_{\nu\quad\rho )}^{\alpha\quad\beta}=-{D-3\over (D-2)(D+2)} g_{(\mu\nu}Z^{\alpha\beta}_{\quad\rho)},\,.
\een
It would be interesting to have examples of such spacetimes, if they exist at all. If they do, the simplest consistent solution of (\ref{cona},\ref{conaa}) in such a geometry would correspond to setting $a_7=b_3=0$. We would then be left to specify the background field $S_\rho$ (since $U_{\rho\lambda}$ is completely determined from $S_\rho$ and the ambient spacetime).

\item When on the other hand, $Y_{\mu\nu\rho}=Z_{\m\n\p}=0$, then according to (\ref{XYZdef})
\eqn
\nabla^\mu \CC_{\nu\quad\mu}^{\quad (\rho\quad\sigma )}=0\,.
\een
In this case, a simple solution corresponds to setting $a_2=b_1=0$, together with
\eqn\label{solYZzero}
a_7 W_{\m\n\p\s}=-s(s-1)\,\CC_{\m\n\p\s} +\Lambda_{\m\n\p\s}\,,
\een
where $\Lambda_{\m\n\p\s}$ is an arbitrary tensor with the symmetry properties of the Weyl tensor and additionally satisfies:
\eqn\label{Lambdacon}
\nabla_{(\kappa} \Lambda_{\m \quad \n)}^{\quad \p\quad \s}=0,\qquad g^{\m\p}\Lambda_{\m\n\p\s}=g^{\n\s}\Lambda_{\m\n\p\s}=0 \,.
\een
For instance, $\Lambda_{\m\n\p\s}$ can be equal to the Weyl tensor of any symmetric space.

One should then determine $\hS_{\m |\p\s}$ by combining (\ref{conaasb}) together with (\ref{conaa}), ensuring in particular that:
\eqn
{a_7\over s} W_{(\p |\m |\s)\n}= b_3\nb_{\m}\hS_{\n\p\s}\,. 
\een

A large number of spacetimes belongs to this class. Recall that the divergence of the Weyl tensor can be expressed in terms of the Cotton tensor:
\eqn
\nb^\s \CC_{\m\n\p\s}=(D-3) C_{\m\n\p} 
\een
which is in turn defined in terms of the Schoutten tensor through
\eqn
\begin{split}
C_{\mu\nu\rho}\equiv{2\over D-2}\nb_{[\m} Sch_{\n]\p},\qquad\text{where}\qquad (Sch)_{\p\s}\equiv R_{\p\s}-{1\over 2(D-1)} \,g_{\p\s} R\,.
\end{split}
\een
Clearly, geometries with vanishing Cotton tensor automatically satisfy $Z_{\mu\nu\rho}=0$. Ricci flat metrics, are the simplest examples with $Y_{\mu\nu\rho}=Z_{\m\n\p}=0$. Interesting spacetimes in this class are the asymptotically flat Schwarzchild and Kerr black hole solutions. Asymptotically dS or AdS Einstein metrics also belong into the class of spacetimes which satisfy $Y_{\m\n\p}=Z_{\m\n\p}=0$. Specific examples are the asymptotically (A)dS-Schwarzchild and (A)dS-Kerr black holes \cite{Aminneborg:1996iz, Carter:1968ks, Carter:1968rr, Hawking:1998kw}. 
\end{itemize}

\end{itemize}

\paragraph{Fields of spin $s=2$.}

\begin{itemize}

\item The simplest possible solution to the set of conditions (\ref{conb}) corresponds to setting $a_1=b_2=a_2=0$, and reducing the system to:
\eqn\label{eqnfinalstwosimple}
\begin{split}
&\left[\nb^2-m^2+\alpha R\right]\vf_{\m_1...\m_s}+2 R_{(\m_1}{}^{\n}{}_{\m_2}{}^{\p}\vf_{\m_3\dots\m_s)\n\p}-2 R_{\p(\m_1}\vf^{\p}{}_{\m_2\dots\m_s)}=0,\cr
&\nb\cdot\varphi_{\m_1\dots\m_{s-1}}+b_2 S_{\rho\sigma}^{\,\,\,(\mu_1}\vf^{\mu_2\cdots\mu_{s-1})\rho\sigma}+b_3 \hS^{(\m_1}_{\,\,\,\p\s}\vf^{\,\m_2\cdots\m_{s-1})\p\s}=0,\qquad \vf^{\mu(s-2)}=0
\end{split}
\een
with 
\eqn\label{conaabs}
b_2\nabla^2 S_{\nu\rho\mu}+b_3\nabla^2 \hS_{\mu|\nu\rho}=Y_{\mu\nu\rho}-{2\over 3} Z_{\nu\rho\mu}\quad \text{and}\quad \alpha={2D\over (D-1)(D+2)}\,.
\een
Note that in this case only the constraint equation is modified and the relevant background fields satisfy a rather simple equation.

\item It is clear that $Y_{\mu\nu\rho}$ does not need to vanish and as a result it is possible to use the involutive system (\ref{eqnfinalstwo}) or (\ref{eqnfinalstwosimple}) to describe the propagation of spin-2 fields in conformally flat spacetimes.
Conformally flat spacetimes are particularly simple to treat because $Z_{\mu\nu\rho}=X_{\mu\nu\rho}^{\quad\mu_1\mu_2}=0$.

\item Particular cases of interest are the asymptotically AdS domain-wall geometries (relevant for holographic applications) and the Friedmann-Robertson-Walker metrics (relevant for cosmology). Consider the following metric:
\eqn\label{cfmetric}
ds^2=g_{\m\n}^{DW}dx^\m dx^\n=dy^2+e^{2 f(y)}\left[-(1-k r^2)dt^2+{dr^2\over 1-k r^2}+r^2d\Omega_{D-3}\right],\qquad -\infty<y<+\infty,
\een
where $D$ denotes the dimensionality of the spacetime and $\Omega_{D-3}$ the unit hypersphere. $k$ takes the values $(-1,0,+1)$ which correspond to $(D-1)$-dimensional AdS, flat and dS slicings respectively. The metric is conformally flat, so $Z_{\mu\nu\rho}=X_{\mu\nu\rho}^{\quad\mu_1\mu_2}=0$ automatically. On the other hand, $Y_{\mu\nu\rho}$ does not vanish but takes a rather simple form with non-zero components:
\eqn\label{cfY} 
\begin{split}
Y_{yyy}&={(D-2)(D-1)\over D+2} e^{-2 f(y)} \,\left[4 k f'(y)+e^{2 f(y)} (2 f'(y) f''(y)-f'''(y))\right] g_{yy},\cr
Y_{yii}=Y_{iyi}=Y_{iiy}&=-{D-2\over D+2} e^{-2 f(y)} \,\left[4 k f'(y)+e^{2 f(y)} (2 f'(y) f''(y)-f'''(y))\right] g_{ii}, \quad\text{for}\quad i=(t,r,\Omega)
\end{split}
\een

In order to find a solution of (\ref{conaabs}) we set $b_3=0$ and make the following ansatz: 
\eqn\label{Sansatz}
b_3\,S_{\m\n\p}^{DW} = {s(y)\over 4 k f'(y)+e^{2 f(y)} (2 f'(y) f''(y)-f'''(y)) } Y_{\m\n\p}^{DW}\,.
\een
Substituting the ansatz into (\ref{conaabs}) leads to a differential equation $s(y)$ must satisfy for consistency:
\eqn\label{seq}
s''(y)+(D-5) f' s'(y)-2\,\left[\left({5(D-1)\over 2}+1\right) f'^2+f''\right]\,s(y)=4 k f' +e^{2 f(y)} (2 f' f''-f'''),
\een
Thus, given a conformally flat metric (\ref{cfmetric}) with known $f(y)$,  the involutive system (\ref{eqnfinalstwosimple}) describing the propagation of spin-2 fields in this background metric is consistent as long as another background field, $S_{\m\n\p}$ , which satisfies eqs (\ref{Sansatz},\,\ref{seq}) is turned on.

For completeness, we present also the case of the four-dimensional Friedmann-Robertson-Walker metric:
\eqn\label{FRWcosmo}
ds^2=g_{\m\n}^{FRW}dx^\m dx^\n=-dt^2+a(t)^2\left[{dr^2\over 1-k r^2}+r^2d\Omega_2\right]
\een
with 
\eqn\label{Yfrw}
\begin{split}
Y_{ttt}&=\left(4+4 k {a'\over a^3}-5 {a' \,a''\over a^2}+{ a'''\over a}\right)\,g_{tt}^{FRW}\cr
Y_{jtj}=Y_{jjt}=Y_{tjj}&={1\over 3}\,\left(4+4 k {a'\over a^3}-5 {a' \,a''\over a^2}+{ a'''\over a}\right)\,g_{jj}^{FRW},\qquad\text{for}\qquad j=(r,\Omega)
\end{split}
\een
It is easy to see that
\eqn
b_3\,S_{\m\n\p}^{FRW}={h(t) \over 4+4 k {a'\over a^3}-5 {a' \,a''\over a^2}+{ a'''\over a} }\,\,Y_{\m\n\p}^{FRW}
\een
as long as $h(t)$ solves the following differential equation:
\eqn
h''(t)+{3 a'\over a} h'(t)-15 {a'^2\over a^2}h(t)=-\left(4+4 k {a'\over a^3}-5 {a' \,a''\over a^2}+{ a'''\over a}\right)
\een

\end{itemize}

\section{Conditions for an exact solution}

\noindent In the previous sections we restricted our attention to the cancellation of the anomaly tensor to first order in the curvature. To obtain the cancelation without imposing restrictions on the ambient spacetime, we introduced new background fields. With the help of these new fields, our goal was reached. However, the new fields are bound to modify the anomaly tensor to second order in the curvature. This definitely limits the applications of our results. 

For an all order's result, we must require the cancellation of the anomaly to second order in the curvature (higher order cancellation will be automatically achieved).  
To this end, we compute here the $\OO(R^2)$ and demand that they vanish, {\it i.e.}
\eqn
(M_1\triangleright T_1)^{\alpha(s-1)}+(M_1\triangleright U)^{\alpha(s-1)}+(L_1\triangleright U)^{\alpha(s-1)}=0\,.
\een
The computation is starightforward given the formuli for $L_1, M_1$ given in eqs (\ref{Lonedef}) and (\ref{M11})-(\ref{M12}).

\paragraph{Fields of spin $s\geq 3$.}

\noindent Explicit computation yields:
\eqn\label{OOtwo}
\begin{split}
&b_1 S_\lambda \left[\left(-R^{\rho\lambda}+\frac{a_2}{s} U^{\rho\lambda}\right)\varphi_\rho^{\,\,\alpha_1\cdots\alpha_{s-1}}-2\left((s-1)R^{\rho\lambda\sigma(\alpha_1}+\frac{a_7}{s}W^{\rho\lambda\sigma(\alpha_1} \right)\varphi^{\alpha_2\cdots\alpha_{s-1})}_{\qquad\qquad\rho\sigma}        \right]+\cr
&+b_3\left[\hS^{\rho |}_{\quad\kappa\lambda} \left(R_\rho ^{\,\,\,(\alpha_1} -\frac{a_2}{s} U_\rho^{\,\,\,(\alpha_1}\right)\varphi^{\alpha_2\cdots\alpha_{s-1})\kappa\lambda}-2 \left(R_\rho^{\,\,\,\kappa}-\frac{a_2}{s}U_\rho^{\,\,\,\kappa}\right)\hS^{(\alpha_1 |}_{\qquad\kappa\lambda}\varphi^{\alpha_2\cdots\alpha_{s-1})\rho\lambda}\right]+\cr
&+2 b_3\left(R_{\rho\,\,\,\sigma}^{\,\,\kappa\,\,\,\lambda}+\frac{a_7}{s(s-1)} W_{\rho\,\,\,\sigma}^{\,\,\kappa\,\,\,\lambda}\right) \hS^{(\alpha_1 |}_{\qquad\kappa\lambda}\varphi^{\alpha_2\cdots\alpha_{s-1})\rho\sigma} -\cr
&-2 b_3 \hS^{\rho |}_{\quad\kappa\lambda}\left(R_{\rho\quad\,\sigma}^{\,\,\,(\alpha_1\quad\alpha_2}+\frac{a_7}{s(s-1)} W_{\rho\quad\sigma}^{\,\,(\alpha_1\,\,\,\alpha_2}\right)\varphi^{\alpha_3\cdots\alpha_{s-1})\sigma\kappa\lambda}+\cr
&+4(s-2) b_3  \left(  R_{\rho\,\,\,\sigma}^{\,\,\,\,\kappa\,\,\,(\alpha_1}+\frac{a_7}{s(s-1)} W_{\rho\,\,\,\sigma}^{\,\,\kappa\,\,\,(\alpha_1}  \right)\hS^{\alpha_2 |}_{\quad\kappa\lambda}\varphi^{\alpha_3\cdots\alpha_{s-1})\lambda\rho\sigma}=0
\end{split}
\een
where we took into account that $a_3=a_6=b_2=a_4-a_5=0$. Eq. (\ref{OOtwo}) leads to the following necessary conditions the background fields should satisfy:
\eqn\label{conaatwo}
\begin{split}
&\quad \,\,\,b_1 S_\lambda \left(-R^{\rho\lambda}+\frac{a_2}{s} U^{\rho\lambda}\right)=0\cr
&-2 b_1 S^\lambda \left( (s-1)R^{(\rho\quad\sigma)\alpha_1}_{\quad\lambda}+\frac{a_7}{s}W^{(\rho\quad\sigma)\alpha_1}_{\quad\lambda} \right) +b_3 \hS_{\lambda |}^{\quad\rho\sigma} \left(R^{\lambda\alpha_1} -\frac{a_2}{s} U^{\lambda\alpha_1}\right) -\cr
&-2 b_3 \left(R^{(\rho}_{\quad\lambda}-\frac{a_2}{s}U^{(\rho}_{\quad\lambda}\right)\hS^{\alpha_1|\lambda |\sigma)}+ 2 b_3\left(R^{(\rho\,\,\,\sigma )}_{\,\,\kappa\,\,\,\lambda}+\frac{a_7}{s(s-1)} W^{(\rho\,\,\,\sigma )}_{\,\,\kappa\,\,\,\lambda}\right) \hS^{\alpha_1|\kappa\lambda} =0\cr
&-2 b_3 \hS^\kappa_{\quad\rho\lambda}\left(R_{\kappa\,\,\,\sigma}^{\,\,(\alpha_1\,\,\,\alpha_2)}+\frac{a_7}{s(s-1)} W_{\kappa\,\,\,\sigma}^{\,\,(\alpha_1\,\,\,\alpha_2)}\right)+4(s-2) b_3  \left(  R_{\rho\,\,\,\sigma}^{\,\,\kappa\,\,\,(\alpha_1}+\frac{a_7}{s(s-1)} W_{\rho\,\,\,\sigma}^{\,\,\kappa\,\,\,((\alpha_1}  \right)\hS^{\alpha_2)}_{\quad\kappa\lambda}=0
\end{split}
\een
The first line of (\ref{conaatwo}) leads to the following simple solutions:
\eqn
a_2=0 \quad\text{or}\quad S_\lambda\nabla^\rho S^\lambda=0\quad \Leftrightarrow\quad S^2=0\,,
\een
where we used (\ref{conaasa}) to set $S_\lambda R^{\lambda\sigma}=0$ and (\ref{conaa}) to express $U_{\rho\sigma}$ in terms of $S_\sigma$.
The other two equations in (\ref{conaatwo}) do not yield a simple solution but for specific curved spacetimes they can be combined with (\ref{conaasb}) and (\ref{conaa})
to check whether an all orders solution exists. 

\paragraph{Fields of spin $s=2$.}

The conditions for spin-two fields can be worked out in the same manner as the general case. Focusing on the simple solution where the only non-vanishing parameters are $(b_2,\,b_3)$ we obtain the following quadratic equations:\eqn\label{secondorderspintwo}
\begin{split}
&b_2\left[S^{\lambda\rho\sigma}R_\lambda^{\quad a_1}-2 S^{a_1\lambda (\rho}R^{\sigma )}_{\quad \lambda}+2 S^{a_1\nu\lambda}R^{(\rho\quad\sigma)}_{\quad\nu\quad\lambda}\right]=0\cr
&b_3\left[ \hS^{\lambda |\p\s}R_\lambda^{\quad a_1}-2 \hS^{a_1|\lambda (\p}R^{\quad \s )}_\lambda  +2 \hS^{a_1|\kappa\lambda}R^{(\p\quad\s )}_{\quad\kappa\quad\lambda}    \right]=0
\end{split}
\een
A comment is in order. Eqs (\ref{secondorderspintwo}) are sufficient but not necessary conditions for the existence of an all orders solution. The reason is that for spin-2 fields, various background fields can be turned on without any effect on the anomaly to linear order in the curvature. Such background fields will in general contribute to the anomaly only to second order in the curvature. One may thus fully determine them from the requirement of an all-orders cancellation of the anomaly, whenever (\ref{secondorderspintwo}) is not satified.

\section{Discussion \& Outlook}

\noindent In this article we used the involutive method to derive equations describing the consistent propagation of higher spin fields in non-trivial spacetimes. To expand the set of possible background geometries where consistent propagation is possible, we allowed for additional background fields to be turned on. We remained agnostic as to the nature of these background fields. For simplicity, we limited the discussion to deformations of the dynamical equations and divergence constraints, leaving the trace constraint undeformed. We found that massive fields of spin $s\geq 3$ cannot propagate consistently on geometries characterised by $Y_{\m\n\p}=0$. These unfortunately include all the non-trivial domain-wall type metrics. We were however able to show that consistent propagation may be possible in a large set of non-trivial spacetimes, such as asymptotically flat or dS/AdS black hole geometries, as long as certain conditions between the various background fields are satisfied. On the other hand, massive spin-2 fields are found to propagate consistently on dS, flat or AdS-domain wall spacetimes. Section 4 contains the main results of this work. 

Several open questions remain. Most of them are related to the explicit or implicit assumptions under which this work was carried out. For reasons of completeness we list them here:
\begin{itemize}
\item Consistency of free propagation of massive higher spin fields in arbitrary backgrounds is related to the existence of an abelian algera defined by the d'Alembertian, the divergence and the trace operators. 

This assumption is plausible, being a natural generalisation of the flat space case. Nonetheless, other possibilities may also exist. Indeed, the authors of  \cite{Rahman:2016tqc} argued that a bigger non-abelian algebra, which includes the algebra discussed here, may be necessary for consistency. It would be interesting to include other background fields in the analysis of  \cite{Rahman:2016tqc} and examine the implications.

\item The trace constraint remains undeformed.

In other words, the trace vanishes identically instead of as an on-shell condtion. The assumption was made for the technical simplification it provides. It is clear that it may have non-trivial consequences in a possible Lagrangian formulation of the system. On the other hand, as discussed in \cite{Kulaxizi:2014yxa}, one may view the original system (\ref{fpflatzero}) as the zero-trace gauge fixing of a system of symmetric rank-s fieds with Weyl symmetry: $\delta\vf_{a_1\cdots a_s}=g_{(a_1 a_2}\lambda_{a_3\cdots a_s)}$. Then the freedom of the rank-$(s-2)$ parameter $\lambda_{a_1\cdots a_{s-2}}$ allows one to choose the trace to vanish at the interaction level.  

One may wonder whether, had we allowed for the deformation of the trace constraint, our results would be modified. In particular, would backgrounds with $Y_{\m\n\p}\neq 0$ be allowed? We leave this as future work.

\item At the linearised level, higher spin fields decouple from each other as well as from background fluctuations. 

In practice we assumed that a basis exists such that diagonalisation is possible. This is a natural assumption when considering the free propagation of massive higher spin fields in curved spacetime, like in \cite{Cortese:2013lda, Kulaxizi:2014yxa}. However, one expects mixing to occur at least with graviton fluctuations, when other background fields are present. For the sake of simplicity, we did not consider this case here but is definitely worth of further investigation.

\item Inhomogeneous terms included in the dynamical equations and constraints need not be considered.

This assumption is justified from the standard analysis of linearised fluctuations in Lagrangian systems. 
At the same time, inhomogeneous terms would change the nature of the algebra formed by the d'Alembertian, the divergence and the trace operators. As a result, they would appear to spoil the consistency of the system. It is likely however, that with a careful choice of such terms, we could preserve the algebraic structure but promote it to a non-abelian one. In \cite{Rahman:2016tqc}, a bigger non-Abelian algebra was considered as necessary for consistency. In this spirit, the requirement of an abelian algebra may be too strong. This point certainly deserves further study.

\item Fields of mixed symmetry need not be included.

The absence of mixed-symmetry fields may be too strong of an assumption. It appears that in string theory neglecting mixed-symmetry fields corresponds to considering only the first Regge trajectory of string excitations, thereby excluding all the subleading ones \cite{Rahman:2016tqc}. Put differently, consistency in string theory requires the inclusion of mixed-symmetry fields. In a positive scenario, our simplified approach may lead to possibilities which are not realised in string theory. In a negative one, it may strongly restrict the set of backgrounds where consistency of propagation can be achieved.

\item Higher-derivative, including various other quadratic terms, need not be included.

By preserving the derivative structure of the system, we ensure hyperbolicity and avoid superluminality. 
However, it is indeed expected that higher-derivative terms may be necessary for the contstruction of a consistent theory. At the linear level it may not be so crucial, nevertheless it would be interesting to see if more general spacetimes are allowed if this assumption is removed.

\end{itemize}

We would also like to draw our attention to the treatment of background fields. We did not specify the precise nature of the various tensors appearing in (\ref{Utensor}) and (\ref{Stensor}). They may correspond to higher spin fields taking on background values or they may be composed from expectation values of low spin fields and the metric tensor, similarly to the spacetime tensors $X_{\m\n\p}^{\qquad\s\lambda},Y_{\m\n\p},Z_{\m\n\p}$. However, we assumed that whichever background fields are turned on, they do not alter the structure of the covariant derivative. This is a very strong assumption which may significantly restrict the applicability of our results.
Moreover, we did not search for a theory which could derive the equations of section 4, involving various background fields and gravity. It would be interesting to see if such a theory can be constructed and/or if it is related to string theory in any way. One possibility, would be to consider string solutions supported on a non-trivial dilaton and a closed two-from $B_{\mu\nu}$.

It is a curious fact that, for spacetimes whose Ricci tensor is covarianty constant, the involutive system obtained in \cite{Cortese:2013lda} coincides for spin $s=2$ fields with the system of equations derived to linear order in ($\alpha'$) from the string sigma model \cite{Buchbinder:1999be}. It would be interesting to see if the matching persists when additional string theory background fields are considered.
Note however that in the presence of supersymmetry our results are likely to be modified.

Finally, we would like to mention posible applications of our results. 
Immediate applications concern the study of spin $s=2$ fields in AdS-domain wall geometries both at zero and finite temperature. Similar studies in FRW cosmologies would also be of interest. In practice, one would like to study the equations semiclassically and determine whether bound or quasi-bound states would be possible. A stability analysis should also be performed.  

Recently, there has been renewed interest in the study of massive spin $s=2$ fields, motivated by the contruction of a consistent massive (bi)gravity theory \cite{deRham:2010ik, deRham:2010kj}. Here we did not attempt to construct a full-fledged theory of higher spin massive particles, but focused only on their propagation in external backgrounds.  A hasty calculation shows that the linearised equations of  \cite{deRham:2010ik, deRham:2010kj} differ not only from the ones obtained herein but also from \cite{Cortese:2013lda, Kulaxizi:2014yxa} (see for example, \cite{Bernardetal, Bernard:2015uic} or \cite{Brito:2016peb} and references therein). It would be interesting to confirm this and study its phenomenological implications.

For higher spin fields, the simplest and most interesting case involves black hole geometries. In particular, investigations in the context of the AdS/CFT correspondence may be relevant for the description of different phases in strogly coupled quantum field theories. 


\section*{Acknowledgements}
It is a pleasure to thank R. Rahman for many useful discussions and a critical reading of the manuscript.
I.C. acknowledges support in part from CONACyT (Mexico) grant 238734 and DGAPA-UNAM (Mexico) grant IN107115. M.K. acknowledges support in part from the Hamilton Mathematics Institute and the Simons Foundation.

\begin{appendix}
\numberwithin{equation}{section}

\section{Identities}
\noindent Here we prove the identity
\eqn
\hV_{(\rho|\sigma\nu)\mu}+\hV_{\mu|\rho\sigma\nu}=-2 \hV_{(\rho|\sigma\nu)\mu}
\een
starting from the total symmetrization of $\hV_{\rho|\sigma\nu\mu}$
\eqn
0=\hV_{(\rho|\sigma\nu\mu)}={1\over 4} \left( \hV_{\rho|\sigma\nu\mu}+\hV_{\sigma|\nu\mu\rho}+\hV_{\nu|\mu\rho\sigma}+\hV_{\mu|\rho\sigma\nu}\right)={1\over 4} \left(3 \hV_{(\rho|\sigma\nu)\mu}+\hV_{\mu|\rho\sigma\nu}\right)\,.
\een
Notice also that:
\eqn
3 \hV_{(\rho|\sigma\nu)\mu}+\hV_{\mu|\rho\sigma\nu}=2 \hV_{(\rho|\sigma\nu)\mu} +\hV_{(\rho|\sigma\nu)\mu} +\hV_{\mu|\rho\sigma\nu}=2 (\hV_{(\p | \s) \nu\mu}+\hV_{(\nu |\nu)\p\s})
\een
leading to:
\eqn\label{corrolaryid}
\hV_{(\p |\s )\nu\mu}=-\hV_{(\nu | \mu)\p\s}
\een

\section{Equations for the background from the linear order analysis}

Here we write explicitly the results from the operations involved in the $(L_0 \triangleright Q_1)^{a(s-1)}$. We find that:
\eqn\label{exr1}
\nabla_\nu (U \vf^{\mu(s-1)\nu})=(\nabla_\nu U)\vf^{\mu(s-1)\nu}+ U \stackrel{T_0^{\mu(s-1)}} {\overbrace{\nabla\cdot\vf^{\mu(s-1)}}}
\een

\eqn\label{exr2}
\begin{split}
\nabla_\nu(U_\rho^{\quad (\mu_1}\vf^{\mu_2\cdots\mu_{s-1}\nu )\rho})&={1\over s}\left[(-\nabla\cdot U_\rho)\vf^{\mu(s-1)\rho}-U_{\nu\rho}\nabla^{[\nu}\vf^{\p ]\mu(s-1)}+\right. \cr
&\qquad\left. +(s-1)\left(\nabla_{(\nu}U_{\p )}^{\quad (\mu_1}\vf^{\mu_s\cdots\mu_{s-1})\nu\p}+U_\p^{\quad (\mu_1}      \stackrel{T_0^{\mu(s-1)}} {\overbrace{  \nabla\cdot\vf^{\mu_s}\cdot\mu_{s-1})\p }} \right)\right]
\end{split}
\een

\eqn\label{exr3}
\begin{split}
\nabla_\nu (V_\p^{\quad (\mu_1}\vf^{\mu_2\cdots\mu_{s-1} \nu )\p})&={1\over s} \left(    (\nabla\cdot V_\p) \vf^{\mu(s-1)\p}  +V_{\nu\p}\nabla^{(\nu}\vf^{\p )\mu(s-1)}\right)+\cr
&+{s-1\over s} \left(   \nabla_{(\nu}V_{\p )}^{\quad (\mu_1}\vf^{\mu_2\cdots\mu_{s-1})\nu\p}+\nabla_\p^{\quad (\mu_1}\stackrel{T_0^{\mu(s-1)}} {\overbrace{   \nabla\cdot\vf^{\mu_2\cdots\nu_{s-1})\p} }}  \right) \cr
\nabla_\nu (V_{\p\sigma} g^{(\mu_1\mu_2}\vf^{\mu_3\cdots\mu_{s-1}\nu )\p\sigma})&={2\over s}  \left( (\nabla^{(\mu_1}V_{\p\sigma}) \vf^{\mu_2\cdots\mu_{s-1})\p\sigma}+ V_{\p\sigma} \nabla^{(\mu_1}\vf^{\mu_2\cdots\mu_{s-1})\p\sigma}\right)+\cr
&+ {s-2\over s} \left(\nabla_{(\nu}V_{\p\sigma )}g^{(\mu_1\mu_2}\vf^{\mu_3\cdots\mu_{s-1})\nu\p} +V_{\p\sigma}g^{( \mu_1\mu_2}     \stackrel{T_0^{\mu(s-1)}} {\overbrace{        \nabla\cdot\vf^{\mu_3\cdots\mu_{s-1})\p\sigma} }}\right)
\end{split}
\een

\eqn\label{exr4}
\begin{split}
&\nabla_\nu (\hV_{\p |\sigma}^{\quad (\mu_1\mu_2}\vf^{\mu_3\cdots\mu_{s-1} \nu )\p\sigma})={2\over s}\left( \nabla^\nu \hV_{\p|\sigma\nu}^{\quad (\mu_1}\vf^{\mu_2\cdots\mu_{s-1})\p\sigma}+\hV_{\p |\sigma}^{\quad\nu (\mu_1}\nabla_\nu\vf^{\mu_2\cdots\mu_{s-1})\p\sigma} \right) + \cr
& \qquad\qquad\qquad\qquad\qquad\qquad+{s-2\over s} \left(  \nabla_{(\nu}\hV_{\p |\sigma )}^{\quad (\mu_1\mu_2}\vf^{\mu_3\cdots\mu_{s-1})\p\sigma\nu}+\hV_{\p |\sigma}^{\quad (\mu_1\mu_2}         \stackrel{T_0^{\mu(s-1)}} {\overbrace{    \nabla\cdot\vf^{\mu_2\cdots\mu_{s-1})\p\sigma}    }} \right)   \cr
&\nabla_\nu (\hV^{(\mu_1 |\mu_2}_{\qquad\quad \p\sigma}  \vf^{\mu_3\cdots\mu_{s-1}\nu )\p\sigma})=-{2\over s}\left( \nabla^\nu \hV_{(\p | \s )\nu}^{\qquad (\mu_1} \vf^{\mu_2\cdots\mu_{s-1})\p\sigma}+\hV_{\p | \s}^{\quad\nu (\mu_1}
) \nabla_{\nu }\vf^{\mu_2\cdots \mu_{s-1})\p\sigma}\right)+ \cr
&\qquad\qquad\qquad\qquad\qquad\quad\quad +{s-2\over s}\left(\nabla_{\nu}\hV^{(\mu_1\mu_2}_{\qquad\quad\p\sigma }\vf^{\mu_3\cdots\mu_{s-1})\p\sigma\nu}+\hV^{(\mu_1|\mu_2}_{\qquad\quad\p\sigma}  \stackrel{T_0^{\mu(s-1)}} {\overbrace{   \nabla\cdot\vf^{\mu_3\cdots\mu_{s-1})\p\sigma}  }}       \right) \cr
&\nabla_\nu (\hV_{\p |\sigma\lambda}^{\qquad (\mu_1} g^{\mu_2\mu_3}\vf^{\mu_4\cdots\mu_{s-1}\nu )\p\sigma\lambda})={1\over s}\left((\nabla^\nu \hV_{\p |\sigma\lambda\nu})g^{(\mu_1\mu_2}\vf^{\mu_3\cdots\mu_{s-1})\p\sigma\lambda}+ \hV^{\p|\sigma\lambda\nu} g^{(\mu_1\mu_2}\nabla_\nu\vf^{\mu_3\cdots\mu_{s-1})}_{\qquad\p\sigma\lambda}\right)+ \cr
&\qquad\qquad\qquad\qquad\qquad\qquad\qquad\quad+{2\over s} \left( \nabla^{(\mu_1}\hV_{\p |\sigma\lambda}^{\qquad\mu_2}\vf^{\mu_3\cdots\mu_{s-1})\p\sigma\lambda}  +\hV_{\p|\sigma\lambda}^{(\mu_1}\nabla^{\mu_2}\vf^{\mu_3\cdots\mu_{s-1})\p\sigma\lambda}   \right)+\cr
&\qquad\qquad\qquad\qquad\qquad+{s-3\over s} \left(\nabla_\nu \hV_{\p|\sigma\lambda}^{\qquad(\mu_1}g^{\mu_2\mu_3}\vf^{\mu_4\cdots\mu_{s-1})\p\sigma\lambda\nu}+\hV_{\p|\sigma\lambda}^{\qquad (\mu_1} g^{\mu_2\mu_3}          \stackrel{T_0^{\mu(s-1)}} {\overbrace{   \nabla\cdot\vf^{\mu_4\cdots\mu_{s-1})\p\sigma\lambda}   }} \right)\cr
&\nabla_\nu (g^{(\mu_1\mu_2}\hV^{\mu_3 |}_{\quad\p\sigma\lambda}\vf^{\mu_4\cdots\mu_{s-1}\nu )\p\sigma\lambda})=
{2\over s} \left( (\nabla^{(\mu_1}\hV^{\mu_2 |}_{\quad\p\sigma\lambda})\vf^{\mu_3\cdots\mu_{s-1})\p\sigma\lambda}+\hV^{(\mu_1}_{\quad\p\sigma\lambda}\nabla^{\mu_2}\vf^{\mu_3\cdots\mu_{s-1})\p\sigma\lambda}        \right) +\cr
&\qquad\qquad\qquad\qquad\qquad\qquad +{1\over s} \left( (\nabla^\nu\hV_{\nu |\p\sigma\lambda}) g^{(\mu_1\mu_2} \vf^{\mu_3\cdots\mu_{s-1})\p\sigma\lambda}+\hV^{\nu|\p\sigma\lambda}g^{(\mu_1\mu_2}\nabla_\nu\vf^{\mu_3\cdots\mu_{s-1})}_{\qquad\qquad\p\sigma\lambda} \right)+\cr
&\qquad\qquad\qquad\qquad\qquad+{s-3\over s}\left(\nabla_{(\nu}\hV^{(\mu_1 |}_{\quad\p\sigma\lambda}g^{\mu_2\mu_3}\vf^{\mu_4\cdots\mu_{s-1})\p\sigma\lambda\nu}+\hV^{(\mu_1 |}_{\qquad\p\sigma\lambda}g^{\mu_2\mu_3}        \stackrel{T_0^{\mu(s-1)}} {\overbrace{  \nabla\cdot\vf^{\mu_4\cdots\mu_{s-1})\p\sigma\lambda}   }} \right)
\end{split}
\een
where we used the symmetry properties of $\hV_{\mu |\nu\p\s}$ together with the identity (\ref{corrolaryid}) to write:
\eqn
\hV_{\nu |\mu_1\p\sigma}+\hV^{\mu_1 |\nu\p\sigma}  = 2 \, \hV_{(\mu_1 |\nu)\p\s}= -2 \,\hV_{(\p | \s),\nu\mu_1}
\een

Similarly,
\eqn\label{exr5}
\begin{split}
\nabla_\nu (V_{\p\s}^{\quad (\mu_1\mu_2}\vf^{\mu_3\cdots\mu_{s-1}\nu )\p\s})&={2\over s} \left(\nabla^\nu V_{\p\s\nu}^{\qquad (\mu_1}\vf^{\mu_2\cdots\mu_{s-1})\p\s}+V^{\p\s\nu(\mu_1}\nabla_\nu\vf^{\mu_2\cdots\mu_{s-1})}_{\qquad\qquad\p\s}\right) +\cr
&+{s-2\over s}\left(     \nabla_{(\nu}V_{\p\s )}^{\quad (\mu_1\mu_2}\vf^{\mu_3\cdots\mu_{s-1})\p\s\nu}+V_{\p\s}^{\quad (\mu_1\mu_2} \stackrel{T_0^{\mu(s-1)}} {\overbrace{  \nabla\cdot\vf^{\mu_3\cdots\mu_{s-1})\p\s} }}  \right)\cr
\nabla_\nu (V_{\p\s\lambda}^{\qquad (\mu_1} g^{\mu_2\mu_3}\vf^{\mu_4\cdots\mu_{s-1})\p\s\lambda})&={1\over s}\left((\nabla^\nu V_\p\s\lambda\nu) g^{(\mu_1\mu_2}\vf^{\mu_3\cdots\mu_{s-1})\p\s\lambda}+V^{\p\s\lambda\nu}g^{(\mu_1\mu_2}   \nabla_\nu\vf^{\mu_2\cdots\mu_{s-1})}_{\qquad\qquad\p\s\lambda}\right)+\cr
&+{2\over s}\left( \nabla^{(\mu_1}V_{\p\s\lambda}^{\qquad\mu_2}\vf^{\mu_3\cdots\mu_{s-1})\p\s\lambda}+V_{\p\s\lambda}^{\qquad (\mu_1}\nabla^{\mu_2}\vf^{\mu_3\cdots\mu_{s-1})\p\s\lambda}\right)+\cr
&+{s-3\over s}\left( \nabla_{(\nu}V_{\p\s\lambda}^{\qquad (\mu_1}g^{\mu_2\mu_3}\vf^{\mu_4\cdots\mu_{s-1})\p\s\lambda\nu}+V_{\p\s\lambda}^{\qquad (\mu_1}g^{\mu_2\mu_3} \stackrel{T_0^{\mu(s-1)}} {\overbrace{    \nabla\cdot \vf^{\mu_4\cdots\mu_{s-1})\p\s\lambda}  }}  \right)\cr
\nabla_\nu (V_{\p\s\lambda\kappa}g^{(\mu_1\mu_2}g^{\mu_3\mu_4}\vf^{\mu_5\cdots\mu_{s-1})\p\s\lambda\kappa})&={4\over s}\left(\nabla^{(\mu_1}V_{\p\s\lambda\kappa}g^{\mu_2\mu_3}\vf^{\mu_4\cdots\mu_{s-1})\p\s\lambda\kappa}+V_{\p\s\lambda\kappa}g^{(\mu_1\mu_2}\nabla^{\mu_3}\vf^{\mu_4\cdots\mu_{s-1})\p\s\lambda\kappa} \right)+\cr
&+{s-4\over s}\left(  \nabla_{(\nu}V_{\p\s\lambda\kappa)}g^{(\mu_1\mu_2}g^{\mu_3\mu_4}\vf^{\mu_5\cdots\mu_{s-1})\p\s\lambda\kappa\nu}+V_{\p\s\lambda\kappa}g^{(\mu_1\mu_2}g^{\mu_3\mu_4}    \stackrel{T_0^{\mu(s-1)}} {\overbrace{  \nabla\cdot\vf^{\mu_5\cdots\mu_{s-1})\p\s\lambda\kappa}   }}  \right)
\end{split}
\een

\eqn\label{exr6}
\begin{split}
\nabla_\nu (W_{\p\quad\s}^{\,\, \,(\mu_1\quad\mu_2}\vf^{\mu_3\cdots\mu_{s-1})\p\s})&={2\over s}\left(\nabla^\nu W_{\p\nu\s}^{\qquad (\mu_1}\vf^{\mu_2\cdots\mu_{s-1})\p\s}+W^{
\p\nu\sigma (\mu_1}\nabla_\nu \vf^{\mu_2\cdots\mu_{s-1})}_{\qquad\qquad\p\s}  \right)+\cr
&+{s-2\over s}\left(\nabla_{(\nu}W_{\p\quad\s}^{\,\,\, (\mu_1\,\,\,\mu_2}\vf^{\mu_3\cdots\mu_{s-1})\p\s\nu}+W_{\p\quad\s}^{\,\,\, (\mu_1\,\,\,\mu_2} \stackrel{T_0^{\mu(s-1)}} {\overbrace{    \nabla\cdot\vf^{\mu_3\cdots\mu_{s-1})\p\s}  }} \right)
\end{split}
\een

And from the constraint:
\eqn\label{exr7}
\begin{split}
&\nabla^2 (S_\p\vf^{\mu(s-1)\p})=(\nabla^2 S_\p)\vf^{\mu(s-1)\p}+2 (\nabla_\nu S_\p)\nabla^\nu\vf^{\mu(s-1)\p}+S_\p  \stackrel{T_0^{\mu(s)}} {\overbrace{       \nabla^2 \vf^{\p\mu(s-1)} }}\cr
&\nabla^2 (S_{\p\s}^{\quad (\mu_1}\vf^{\mu_2\cdots\mu_{s-1})\p\s})=(\nabla^2 S_{\p\s}^{\quad (\mu_1})\vf^{\mu_2\cdots\mu_{s-1})\p\s}+2 (\nabla^\nu S_{\p\s}^{\quad (\mu_1})\nabla_\nu \vf^{\mu_2\cdots\mu_{s-1}) \p\s}+S_{\p\s}^{\quad (\mu_1}    \stackrel{T_0^{\mu(s)}} {\overbrace{  \nabla^2\vf^{\mu_2\cdots\mu_{s-1})\p\s}   }} \cr
&\nabla^2 (S_{\p\s\nu}g^{(\mu_1\mu_2}\vf^{\mu_3\cdots\mu_{s-1})\p\s\nu})=(\nabla^2 S_{\p\s\nu})g^{(\mu_1\mu_2}\vf^{\mu_3\cdots\mu_{s-1})\p\s\nu}+2 (\nabla^\lambda S_{\p\s\nu})g^{(\mu_1\mu_2}\nabla_\lambda\vf^{\mu_3\cdots\mu_{s-1})\p\s\nu}+\cr
&\qquad\qquad\qquad\qquad\qquad\qquad  +S_{\p\s\nu}g^{(\mu_1\mu_2}   \stackrel{T_0^{\mu(s)}} {\overbrace{     \nabla^2\vf^{\mu_3\cdots\mu_{s-1})\p\s\nu}  }} \cr
&\nabla^2 (\hS^{(\mu_1 |}_{\quad\,\, \p\s}\vf^{\mu_2\cdots\mu_{s-1})\p\s})= (\nabla^2\hS^{(\mu_1 |}_{\quad\,\,\p\s})\vf^{\mu_2\cdots\mu_{s-1})\p\s}+2 (\nabla^\lambda\hS^{(\mu_1}_{\quad\,\,\p\s})\nabla_\lambda\vf^{\mu_2\cdots\mu_{s-1})\p\s}+\hS^{(\mu_1 |}_{\quad\,\,\,\p\s} \stackrel{T_0^{\mu(s)}} {\overbrace{    \nabla^2\vf^{\mu_2\cdots\mu_{s-1})\p\s} }}
\end{split}
\een
Gathering the above terms with appropriate coefficients according to (\ref{Utensor}) and (\ref{Stensor}) -- except for those which can be related to a deformation of the constraint and/or the dynamical equations -- defines what is denoted as $\CC^{a(s-1)}$ in section 3. Demanding that $\CC^{a(s-1)}$ cancels out the remaining anomaly terms $\BB^{a(s-1)}$ leads to the following equations between the background fields\footnote{Recall that the cancelation should occur identically, and in particular it should not imply any constraints on the higher spin fields themselves.}:

\eqn\label{e1}
a_1 \nabla_\nu U-{a_2\over s}\nabla^\rho\,U_{\rho\nu}+{a_3\over s}\nabla^\rho V_{\rho\nu}-b_1\nabla^2 S_\nu=0
\een

\eqn\label{e2}
\begin{split}
& a_2{s-1\over s}\,\nabla_{(\nu} U_{\rho )\mu}+\,a_3\,{s-1\over s} \,\nabla_{(\nu}V_{\rho )\mu}-\,2 \,a_3\,{s-1\over s(D-4+2 s)} \,\nabla_\mu V_{\nu\rho}+\cr
&+\,2\,{a_4\over s}\,\nabla^\sigma \hV_{(\nu|\rho)\sigma\mu}+\, {a_5\over s}\nabla^\sigma\left(\hV_{\sigma|\mu\nu\rho}+\, \hV_{\mu|\sigma\nu\rho}\right)+\,2 \, {a_6\over s} \nabla^\sigma V_{\nu\rho\sigma\mu}+\cr
&+\,2\,{a_7\over s} \nabla^\sigma W_{(\nu|\sigma|\rho)\mu}-b_2 \nabla^2 S_{\nu\rho\mu}-b_3\nabla^2\hS_{\mu|\nu\rho}=\cr
&={s-1\over d-2}\left[ (D-6+2 s)\,Y_{\nu\rho\mu}-{2 D-6+s\over 3}\,Z_{\nu\rho\mu}      \right]
\end{split}
\een

\eqn\label{e3}
\begin{split}
&-a_3 {(s-1)(s-2)\over s(D-4+2 s)}\,\nabla_{(\nu}V_{\rho\sigma )}g^{\mu_1\mu_2}+a_4{s-2\over s}\, \nabla_{(\nu} \hV_{\rho |\sigma)}^{\quad\mu_1\mu_2}+\, a_5 {s-2\over s}\,\nabla_{(\nu} \hV^{(\mu_1|\mu_2)}_{\qquad \,\,\rho\sigma )}-\cr
&-{(s-2)(2 a_4+a_5)\over s(D-4+2 s)}\, \left[\nabla^\lambda \hV_{(\rho|\sigma\nu)\lambda}\,\, g^{\mu_1\mu_2}+2 \,\nabla^{(\mu_1}\hV_{(\rho|\sigma \nu)}^{\qquad \mu_2)}\right]-2a_5{s-2\over s(D-4+2 s)}\left( \nabla^{(\mu_1}\hV^{\mu_2)|}_{\qquad \rho\sigma\nu}+{1\over 2}\nabla^\lambda \hV_{\lambda|\rho\sigma\nu}\,\,g^{\mu_1\mu_2}    \right)+\cr
&+a_6 {s-2\over s}\,\nabla_{(\nu} V_{\rho\sigma)}^{\quad\mu_1\mu_2}-\,2\,a_6{s-2\over s(D-4+2 s)}\left(\nabla^\lambda V_{\rho\sigma\nu\lambda}\,\, g^{\mu_1\mu_2}+2\,\nabla^{(\mu_1}V_{\rho\sigma\nu}^{\quad\mu_2 )}\right)+a_7 {s-2\over s}\nabla_{(\nu}W_{\rho\quad \sigma)}^{\,\,\, \mu_1\quad \mu_2}+\cr
&+b_2 {s-2\over D-6+2 s}\nabla^2 S_{\rho\sigma\nu}\,\,g^{\mu_1\mu_2}=-{(s-1)(s-2)\over D-2}\left[(D-2) X_{\nu\rho\sigma}^{\qquad \mu_1\mu_2}+Y_{\nu\rho\sigma}g^{\mu_1\mu_2}\right]
\end{split}
\een

\eqn\label{e4}
{(s-2)(s-3)\over s(D-4+2 s)}\times\left[(2 a_4+a_5)\nabla_{(\nu}\hV_{\rho|\sigma\lambda)}^{\quad\mu}+a_5 \nabla_{(\nu}\hV^{\mu|}_{\quad\rho\sigma\lambda )} +2\,a_6 \left(\nabla_{(\nu}V_{\rho\sigma\lambda)}^{\quad\mu}-{2\over D-6+2 s}\nabla^\mu V_{\nu\rho\sigma\lambda}\right)    \right]=0
\een

\eqn\label{e5}
{(s-2)(s-3)(s-4)\over s(D-4+2 s)(D-6+2 s)}\,a_6 \, \nabla_{(\nu}V_{\rho\sigma\lambda\gamma)}=0
\een

\eqn\label{e6}
-{a_2\over s}U_{\nu\rho}+{a_3\over s}V_{\nu\rho}-2 b_1\nabla_\nu S_\rho=0
\een

\eqn\label{e7}
-2 a_3 {s-1\over s(D-4+2 s)}V_{\rho\sigma}=0
\een

\eqn\label{e8}
2\, {a_4\over s}\hV^{(\rho|\sigma)\nu\mu}+{a_5\over s}\left(\hV^{\nu|\rho\sigma\mu}+\hV^{\mu|\nu\rho\sigma}\right)+2\,{a_6\over s}V^{\rho\sigma\nu\mu}+2{a_7\over s}\, W^{(\rho|\nu|\sigma)\mu}-2 b_2\nabla^\nu S^{\rho\sigma\mu}-2 b_3\nabla^\nu \hS^{\mu|\rho\sigma}=0
\een

\eqn\label{e9}
{s-2\over s(D-4+2 s)}\left[(2a_4+a_5)\hV^{(\rho|\sigma\lambda)\nu}+a_5\hV^{\nu|\rho\sigma\lambda}+2 a_6 V^{\rho\sigma\lambda\nu}-2 b_2 {s(D-4+2 s)\over D-6+2 s}\nabla^\nu S^{\rho\sigma\lambda}\right]=0
\een

\eqn\label{e10}
{s-2\over s(D-4+2 s)}\left[(2 a_4+a_5)\hV_{(\rho|\sigma\nu )}^{\qquad \mu}+a_5 \hV^{\mu |}_{\quad\rho\sigma\nu}+2 a_6 V_{\rho\sigma\nu}^{\quad \mu}    \right]=0
\een

\eqn\label{e11}
a_6 {(s-2)(s-3)\over s(D-4+2 s)(D-6+2 s)}V_{\nu\rho\sigma\lambda}=0
\een

\section{Simplifying the equations for the background fields.}

\noindent We deal here with fields of generic spin $s\geq 3$. The spin $s=2$ case is much simpler and can be similarly treated.
Eqs. (\ref{e7}) and (\ref{e11}) immediately lead to: 
\eqn
a_3=a_6=0.
\een
Setting $a_3=a_6=0$ and requiring that (\ref{e9}) and (\ref{e10}) are compatible with each other leads to:
\eqn
\begin{split}
\left(b_2=0, \quad\text{or}\quad \nabla_\nu S_{\mu\p\s}=0\right),\qquad\text{and}\qquad a_4=a_5,\,\,
\end{split}
\een
where we used the identity (see Appendix B for a proof):
\eqn
\hV_{(\rho |\sigma\nu)\mu}+\hV_{\mu|\rho\sigma\nu}=-2 \hV_{(\rho |\sigma\nu)\mu}
\een
to simplify (\ref{e9} - \ref{e10}). Using this identity, it is easy to see that 
\begin{align}
{1\over 2} \left( \hat{V}_{(\rho|\sigma)\nu\mu}+\hat{V}_{(\nu|\mu)\rho\sigma}\right)&=0\cr
(2 a_4+a_5) \hat{V}_{(\rho|\sigma\nu)\mu}+a_5 \hat{V}_{\mu|\rho\sigma\nu}&=2(a_4-a_5)\hat{V}_{(\rho|\sigma\nu)\mu}\,,
\end{align}
which implies that all the terms proportional to $a_4$ and  $a_5$ in( \ref{Utensor}) are actually proportional to the linear combination $(a_4-a_5)$, and thus vanish for $a_4=a_5$.
The set of equations necessary for the vanishing of the anomaly term, can be then expressed simply as follows:
\begin{align}\label{conditions}
{a_7\over s}\nabla_{(\nu} W_{\rho\quad \sigma )}^{\,\,\, \mu_1\quad \mu_2}=-(s-1) X_{\nu\rho\sigma}^{\qquad \mu_1\mu_2}, &\qquad\left(\Rightarrow\quad Y_{\mu\nu\rho}=0,\quad \nabla^\rho W_{\rho\quad \sigma}^{\,\,(\mu_1\quad \mu_2)}=0\right)\cr
{a_7\over s} W^{(\rho\quad \sigma)}_{\quad \nu\quad \mu}-b_3\nabla_\nu \hS_{\mu |}^{\quad \rho\sigma}=0,& \qquad\left(\Rightarrow\quad \nabla^\nu\hS_{\nu|\rho\sigma}=0,\quad \nabla^2\hS_{\mu|\rho\sigma}=0\right)\cr
{a_2\over s} \nabla_{(\nu}U_{\rho)\mu}=-{(2D-6+s)\over 3(D-2)}\,Z_{\nu\rho\mu},&\qquad \left(\Rightarrow\quad\nabla^\rho U_{\rho\mu}=0\right)\cr
{a_2\over s}U_{\nu\rho}+2 b_1\nabla_\nu S_\rho=0,&\qquad\left(\Rightarrow \quad\nabla_{(\nu}S_{\rho )=0},\quad \nabla^\rho S_\rho=0,\quad \nabla^2 S_\rho=0\right)\cr
a_1\nabla_\nu U=0\,.&
\end{align}
Notice, that there is an ambiguity in the value of $U$, which reflects the ambiguity in the definition of the mass parameter in arbitrary spacetimes.

\section{First order deformations of the gauge identity generators.}

\noindent Here we write down explicit expressions for the correction terms to the gauge identity generators. They play an important role for deformations to second (and higher) order in the curvature.
It is straightfowrard to write down $M_{1\qquad\mu (s-1)}^{a(s-1)}$ and $M_{1\qquad\mu (s)}^{a(s-1)}$ using eqs (\ref{exr1}-\ref{exr7}) while bearing in mind the relevant deformation of the dynamical equations and constraints, {\it e.g.}, (\ref{Utensor}) and (\ref{Stensor}). In particular, we obtain:
\eqn\label{M11}
M_{1\qquad\mu (s-1)}^{a(s-1)}=-a_1 U \,\delta^{a_1\cdots a_{s-1} }_{( \mu_1\cdots\mu_{s-1})} -a_2\, {s-1\over s} \,\delta^{a_1\cdots a_{s-1} }_{\p ( \mu_1\cdots\mu_{s-2}} \, U^{\p}_{\mu_{s-1} )} -a_7\,{s-2\over s}  \,  W^{\p\quad\s}_{\,\,\,\, ( \mu_1 \quad\mu_2}\, \delta^{a_1\cdots a_{s-1}}_{\mu_3\cdots\mu_{s-1})\p\s}
\een
and
\eqn\label{M12}
M_{1\qquad\mu (s)}^{a(s-1)}=b_1 S_{( \mu_1} \delta^{(a_1\cdots a_{s-1} )}_{\mu_2\cdots \mu_{s}) }+b_3 \hS^{( a_1 |}_{\quad ( \mu_{1}\mu_2} \delta^{a_2\cdots a_{s-1} )}_{\mu_3\cdots\mu_{s} ) }
\een
where we took into account that $a_3=a_4-a_5=a_6=b_2=0$ as required by the anomaly cancellation to linear order in the curvature for massive fields of spin $s\geq 3$.

In the case of spin $s=2$ massive fields, and for the minimal solution leading to (\ref{eqnfinalstwosimple}), we have that:
\eqn\label{Mspintwo}
\begin{split}
M_{1\quad\mu_1 }^{\,\,a_1}&= 0\cr
M_{1\quad\mu_1\mu_2}^{\,\, a_1}&=b_2 \,S^{a_1}_{\quad\,\,\p\s}\,\delta^{\p\s}_{( \mu_1\mu_2 )}+b_3 \,\hS^{a_1 |}_{\quad\,\, \p\s}\, \delta^{\p\s}_{ ( \mu_1\mu_2 )}
\end{split}
\een

\end{appendix}



\end{document}